\documentclass[11pt]{article}
\usepackage{a4wide}
\usepackage{authblk}
\usepackage{graphicx}

\usepackage{caption}
\usepackage{subcaption}

\usepackage{color}
\usepackage{amsmath,amsfonts,amssymb,graphicx,hyperref,hypcap}
\usepackage{epsfig}
\usepackage{amsmath,amsfonts,epsf}
\usepackage{amssymb,tikz}

\begin{document}
\begin{center}
{\Large\bf  Implications of   JLA data for 
 $k-$essence model of dark energy with  given equation
of state}
\end{center}
\vspace{4mm}
\begin{center}
{\large Abhijit Bandyopadhyay\footnote{Email: abhijit@rkmvu.ac.in} and Anirban Chatterjee\footnote{Email: anirban.chatterjee@rkmvu.ac.in}}
\end{center}

\begin{center}
Department of Physics\\Ramakrishna Mission Vivekananda Educational and Research Institute\\
Belur Math, Howrah 711202, India
\end{center}
\vspace{4mm}

\begin{abstract}

 We investigated implications of recently 
released `Joint Light-curve Analysis' (JLA) supernova Ia (SNe Ia) data
for  dark energy models with  time varying equation of 
state   of dark energy, usually expressed as $w(z)$ in terms
of variation with corresponding redshift $z$. From a comprehensive
analysis of  the JLA data, we obtain the observational constraints on
the different functional forms of $w(z)$, corresponding 
to different varying dark energy models often considered in literature,
\textit{viz.} CPL, JBP, BA and Logarithmic models.
The constraints are expressed in terms of parameters ($w_a, w_b$) 
appearing in the chosen functional form  for $w(z)$, corresponding
to each of the above mentioned models.
Realising dark energy with varying equation of state  in terms
of a homogeneous scalar field $\phi$, with its dynamics
driven by a $k-$essence Lagrangian $L=VF(X)$
with a constant potential $V$  and a dynamical term $F(X)$ with $X=(1/2)\nabla^\mu\phi\nabla_\mu\phi$ we reconstructed 
form of the function $F(X)$.
This reconstruction has been performed for different varying
dark energy models at best-fit values of parameters ($w_a, w_b$)
obtained from analysis of JLA data. In the context of
$k-$essence model, we also investigate the variation of 
adiabatic sound speed squared,
$c_s^2(z)$, and obtained the domains in  ($w_a, w_b$)
parameter space  corresponding to the physical bound $ c_s^2>0$
implying stability of density perturbations.

\end{abstract}

\section{Introduction}
\label{sec:intro}
 
Measurement  of redshift and luminosity distances for 
Type Ia Supernova (SNe Ia) events \cite{Riess98}, \cite{Perlmutter},
are instrumental in establishing the fact that the universe has 
undergone a transition from a phase of decelerated expansion to
accelerated expansion during its late time phase of evolution.
Other independent evidences in support of this fact come  from the 
observations of Baryon Acoustic Oscillation (\cite{Eisenstein},\cite{Zhang},\cite{Aubourg},\cite{Blake},\cite{Seo}), 
 Cosmic Microwave Background radiations 
(\cite{ref:cmb1},\cite{ref:cmb2},\cite{Tristram},\cite{Gawiser})
 measurement of  differential ages 
of the galaxies in GDDS, SPICES and VDSS surveys 
\cite{ref:Riess,ref:simon1,ref:cabre1,ref:stern1} and studies of 
power spectrum of matter distributions of the universe. A general label 
attributed to the origin of this late time cosmic acceleration
is `Dark Energy' (DE).  Besides, study of rotation curves of spiral 
galaxies \cite{Sofue:2000jx}, 
Bullet cluster \cite{Clowe:2003tk}, gravitational lensing \cite{Bartelmann}, provide indirect evidence for existence of non-luminous matter in 
present universe. Such `matter', labelled as `Dark Matter' (DM)  manifest 
its existence only through gravitational interactions.
Measurements in satellite borne experiments - WMAP  \cite{Hinshaw:2012aka}
and Planck \cite{Ade:2013zuv} have established that,
at present epoch, dark energy and dark matter comprise around 96\% 
of total energy density of the universe ($\sim$69\% dark energy and $\sim
$27\% dark matter). Rest $\sim$4\% is
contributed by baryonic matter with negligible contribution from
radiations.\\

There exist diverse theoretical approaches aiming construction of 
different models for dark energy  to explain the present day cosmic
acceleration. These include the $\Lambda-$CDM model \cite{weinberg-cdm}
which provides excellent agreement with the cosmological data.
Here `CDM' refers to Cold Dark
Matter content of the universe and $\Lambda$, the cosmological constant,
denotes vacuum energy density. 
Though this model provides a simple phenomenological solution, it 
is  plagued with the problem of large disagreement
between vacuum expectation value of energy momentum
tensor and observed value of dark energy density (fine tuning problem). 
This motivates  investigation of  alternative models of dark energy.
One of the key features of a class of such models, 
called varying dark energy models, is time varying 
equation of state  $w=p/\rho$ ($\rho$ is the
energy density and $p$ the pressure of dark energy) of 
dark energy which is usually expressed in terms of 
variation of $w$ with redshift $z$ 
(in $\Lambda-$CDM model $w=-1$, constant). 
The redshift dependence of the EOS parameter $w(z)$, 
in the context of the 
varying DE models may be constrained from the observational data.    
The starting point  for dealing with 
the issue of variations of the EOS parameter $w(z)$
is to consider diverse functional forms of $w(z;w_a,w_b)$ 
each involving a small number of parameters
(denoted in the text by symbols $w_a , w_b$). 
The observational constraints on the $z-$dependence of $w$ 
may then be realised in terms of constraints on 
the parameters ($w_a ,w_b$)
for each different functional forms of $w(z;w_a,w_b)$ considered. 
In this work we have performed a comprehensive analysis of 
 `Joint Light-curve Analysis' (JLA) data
(\cite{ref:Suzuki},\cite{Betoule:2014frx},\cite{Wang:2016bba})
to obtain  constraints on  different functional forms of $w(z)$
often used in literature 
in
the context of varying dark energy models 
 (\cite{CPL1},\cite{CPL2},\cite{CPL3},\cite{CPL4},\cite{JBP1},\cite{JBP2},\cite{Barboza1},\cite{Barboza2},\cite{Sangwan},\cite{Wetterich},\cite{DeFelice},\cite{Wang} and references there in).
As benchmark we have chosen four such models
\textit{viz.} CPL\cite{CPL1}, JBP\cite{JBP1},\cite{JBP2}, BA\cite{Barboza1},\cite{Barboza2} and and Logarithmic model \cite{Sangwan},
and for each model
we   presented the  values 
of the parameters ($w_a, w_b$ ) those fit best the  observational
data from JLA
and also  shown the  regions in this
 parameter space at different confidence limits   allowed from   observational data.  \\

Dark energy with varying equation of state may be
realised theoretically
in terms of dynamics of a scalar field  ($\phi$).  
One class of such scalar field
models, called  `Quintessence', is described in terms of standard canonical Lagrangian of 
the form $L = X - V(\phi)$ where $X = \frac{1}{2} \nabla_{\mu}\phi \nabla_{\mu}\phi$ is the kinetic term.
There also exist alternative class of models involving Lagrangians with 
non-canonical kinetic terms as $L =  V(\phi)F(X)$, where $F(X)$ is a function of
$X$. Such models, called $k-$essence models, have  interesting  phenomenological consequences 
different from those of quintessence models.  Another motivation  for considering 
$k-$essence scalar fields is that  they appear naturally 
in low energy effective string theory. Such theories  with non-canonical kinetic terms
was first proposed by Born and Infeld to get rid of the infinite self-energy of the electrons \cite{ref:borninfeld} 
and were also investigated by Heisenberg in the context of cosmic ray physics \cite{ref:hei1} and 
meson production \cite{ref:hei2}. In this work, we consider dark energy represented in terms 
of a homogeneous scalar field $\phi \equiv \phi(t)$ whose dynamics
is driven by a $k-$essence Lagrangian $L = VF(X)$ 
with a constant potential $V$. 
The constancy of the potential ensures existence of a scaling relation, 
$XF_X^2 = Ca^{-6}$ ($C =$ constant), in a Friedmann-Lemaitre-Robertson-Walker (FLRW) background space-time with scale factor $a$.
We exploited the scaling relation and observational constraints
on the parameters $w_a, w_b$, 
 to reconstruct the forms of the function $F(X)$  
 for the different varying dark energy models.  \\
 
In the context of $k-$essence model, we also investigate the adiabatic sound speed squared ($c_s^2$) \cite{xia8} -
the  quantity relevant for the growth of 
small fluctuations in the background energy density. 
Imaginary value of the sound speed ($c_s^2<0$) implies instability of density perturbations.
Also from causality, it requires the speed of propagation of
density perturbations  not to exceed the speed of light ($c_s^2 < 1$). 
However, it was pointed out in \cite{vk1,vk2,vik1} that in 
$k-$essence theories
superluminal propagation of   
perturbations on classical backgrounds is  admissible  
and no causal paradoxes arise. This implies  
the condition $c_s^2>0$ would be enough
to represent the physical bound in the context of $k-$essence theories.
For each of the varying dark energy models considered here,
we find  $z$-dependence of $c_s^2$ at best-fit values of parameters $w_a,w_b$ 
obtained from analysis of JLA data.
This has been found over the entire redshift range $0<z<1.3$  accessible in SNe Ia observations 
corresponding to the JLA data.  We note that 
at the best-fit, $c_s^2$ is not always within
its physical bound ($c_s^2>0$)  for all values of
$z$  in the above mentioned range.  
For each of the varying DE models, we have found the regions in  
$w_a-w_b$ parameter space for which  the physical bound $c_s^2 > 0$ 
is satisfied   for the entire range of values of $z$ 
in JLA data. The best-fit values of 
parameters ($w_a,w_b$) for each model  obtained from
analysis of the observational data are found to lie outside this domain corresponding to the bound $c_s^2>0$ 
(and also to $0<c_s^2<1$ )
implying that observational data  allow values of
parameters ($w_a,w_b$), for which the physical bound on $ c_s^2 $ is
respected, only at higher confidence limits. 
For example, for BA and Logarithmic models
the values of  parameters ($w_a,w_b$) corresponding to  $0<c_s^2(z) < 1$ for all $z$
is allowed from observational data only beyond $\sim 2\sigma$ confidence limits.
It is allowed only at $3\sigma$ and beyond for CPL model and even at
larger confidence limits for JBP model. 
For each of the models we have found the point in $w_a-w_b$ parameter space
which belongs to the domain for which $0<c_s^2 <1 $ for all $z$ 
and is maximally favoured from observational data.
The form of the $k-$essence Lagrangian density $F(X)$ are also 
reconstructed at these points. \\
 
The paper is organised as follows.
In Sec.\ \ref{sec:analysis} we 
describe the methodology of analysis of the observational data and
 provide a brief description of the different data sets used in our analysis.
In Sec. \ref{sec:kessence} we discussed  different models of dark
energy with varying equation of state and their realisations in terms of   
k−essence scalar field models. In this context we also discussed   
relevance of investigating  variations of  the  
adiabatic sound speed squared.
The methodology of obtaining form of the Lagrangian density
$L = VF(X)$ for $k-$essence models are also described.
In  Sec.\ \ref{sec:results1} we discussed the results on
the variation of $c_s^2$, form of the function $F(X)$ obtained
from the analysis of the data. The conclusions are presented in 
Sec.\ \ref{sec:conclusion}.

\section{Methodology of Analysis of Observational Data}
\label{sec:analysis}
Measurement of luminosity distances and redshift of  type Ia supernovae (SNe Ia)  
are instrumental in probing nature of dark energy. There exist several
systematic and dedicated measurements of SNe Ia events. There are different supernova surveys  
in different domains of redshift ($z$).
High redshift projects ($z \sim 1$) include
Supernova Legacy Survey (SNLS) 
(\cite{ref:Astier},\cite{ref:Sullivan}), the ESSENCE project \cite{ref:Wood-Vasey}, 
the Pan-STARRS survey (\cite{ref:Tonry},\cite{ref:Scolnic},\cite{ref:Rest},\cite{xia4}).
The  SDSS-II supernova surveys (\cite{ref:Frieman},\cite{ref:Kessler},\cite{ref:Sollerman},
\cite{ref:Lampeitl},\cite{ref:Campbell}) probe  the redshift regime $0.05<z<0.4$.
The surveys in the small redshift domain  $(z > 0.1)$ are  the Harvard-Smithsonian Center for Astrophysics survey   
\cite{ref:Hicken}, 
the Carnegie Supernova Project  (\cite{ref:Contreras},\cite{ref:Folatelli},\cite{ref:Stritzinger}) 
the Lick Observatory Supernova Search  
\cite{ref:Ganeshalingam}  and the Nearby Supernova Factory  \cite{ref:Aldering}. 
Other different compilations of Sne Ia data may also be
found in (\cite{snrest},\cite{xia6},\cite{xia10}) and references in \cite{ref:Wood-Vasey}.
Nearly one thousand of SNe Ia events were discovered
in all these surveys.\\

The recently released  ``Joint Light-curve Analysis" (JLA) data 
(\cite{ref:Suzuki},\cite{Betoule:2014frx},\cite{Wang:2016bba})
is a compilation of  several low,  intermediate and high redshift samples  
including data from the full three years of the SDSS survey,
first three seasons of the five-year SNLS survey and  14 very
high redshift $0.7 < z < 1.4$ SNe Ia from space-based observations 
with the HST \cite{ref:Riess}.  
This data set contains 740 spectroscopically confirmed SNe IA events with high-quality light curves.\\

In this section we describe the methodology of analysis of JLA data to obtain
bounds on equation of state parameter $w$ of dark energy. 
There exist diverse statistical techniques for analysis 
of JLA data.  Some of these methods are discussed in detail in
(\cite{Sangwan},\cite{xia1},\cite{xia3},\cite{xia5},\cite{xia9}).
However, we take the 
$\chi^2$ function corresponding to JLA data as  
\cite{Betoule:2014frx,Wang:2016bba}
\begin{eqnarray}
\chi^2_{\rm SN} &=& \sum_{i,j}\left(\mu_{{\rm obs}}^{(i)} - \mu_{\rm th}^{(i)}\right)
 \left(\Sigma^{-1}\right)_{ij}  \left(\mu_{\rm obs}^{(j)} - \mu_{\rm th}^{(j)}\right) 
\label{eq:chisqsn}
\end{eqnarray}
where values of the dummy indices $i,j$ run from 1 to 740 corresponding to
the 740  SNe IA events contained in the JLA data set \cite{Betoule:2014frx}.
$\mu_{\rm th}^{(i)}$ stands for the  theoretical  expression for   distance modulus
in a flat FRW spacetime background  for the $i^{\rm th}$ entry of the JLA data set and is given by
\begin{eqnarray}
\mu_{\rm th}^{(i)}  &=& 5\log\left[\frac{d_L  (z_{\rm hel},z_{\rm CMB} )}{\rm Mpc}\right] + 25
\label{eq:muth}
\end{eqnarray}
where
\begin{eqnarray}
d_L  (z_{\rm hel},z_{\rm CMB} ) = (1 + z_{\rm hel}) r(z_{\rm CMB}) \quad \mbox{with} \quad 
r(z) = cH_0^{-1} \int_0^z \frac{dz^\prime}{E(z^\prime)}  \,. \label{eq:dlrz}
\end{eqnarray} 
$d_L$ is the luminosity distance, $r(z)$ is the comoving distance to an object
corresponding to a redshift $z$. $z_{\rm CMB}$ and $z_{\rm hel}$ are SNe IA redshifts in 
CMB rest frame and in heliocentric frame respectively. $c$ is the speed of light and 
$H_0$ is the value of Hubble parameter at present epoch. The function $E(z)$ in
Eq.\  \eqref{eq:dlrz} is the reduced Hubble parameter given by
\begin{eqnarray}
E(z) &\equiv & \frac{H(z)}{H_0}
= \Bigg{\{} \Omega_r^{(0)}(1+z)^4 + \Omega_m^{(0)}(1+z)^3 
+ \Omega_{de}^{(0)}   \exp\left[3\int_0^z dz^\prime \frac{1+w(z^\prime)}{1+z^\prime}\right]
\Bigg{\}}^{1/2}
\label{eq:ez}
\end{eqnarray}
where $\Omega_r^{(0)}$, $\Omega_m^{(0)}$ and $\Omega_{de}^{(0)}$ are the values of
fractional energy density contributions from radiation, matter and dark energy
respectively  at present epoch. \\ 

$\mu_{\rm obs}^{(i)}$ is the observed value of distance modulus at a 
redshift $z_i$ corresponding to  $i^{\rm th}$ entry of the JLA data set .
This is expressed through the following empirical relation as 
\begin{eqnarray}
\mu_{\rm obs}^{(i)} &=& m_B(z_i) - M_B + \alpha \ X_1(z_i) - \beta \ C(z_i)
\label{eq:muobs}
\end{eqnarray}
where $m_B(z_i)$ is  the observed value of peak magnitude,
$X_1(z_i)$ denotes time stretching of the light-curve and $C(z_i)$
is the supernova `color' at maximum brightness. $M_B$ is the absolute magnitude which we take fixed at
$M=-19$ for our work and $\alpha$, $\beta$ are nuisance parameters. $\Sigma$ is the total
covariant matrix given in terms of statistical and systematic uncertainties
as
\begin{eqnarray}
\Sigma_{ij}
&=& \delta_{ij} \Big{[} ({\sigma^2_{z}})_i 
+ ({\sigma^2_{\rm int}})_i
+ ({\sigma^2_{\rm lensing}})_i
+ ({\sigma^2_{m_B}})_i 
+ \alpha^2({\sigma^2_{X_1}})_i 
+ \beta^2({\sigma^2_{C}})_i \nonumber\\
&& + 2\alpha ({\Sigma_{m_B,X_1}})_i
- 2\beta ({\Sigma_{m_B,C}})_i
- 2\alpha\beta ({\Sigma_{X_1,C}})_i
\Big{]}\nonumber \\
&& +  \Big{[}
(V_0)_{ij} + \alpha^2 (V_a)_{ij} + \beta^2 (V_b)_{ij}
+ 2\alpha (V_{0a})_{ij} -  2\beta (V_{0b})_{ij} -   
2\alpha\beta (V_{ab})_{ij}
 \Big{]} 
\label{eq:covariance}
\end{eqnarray}
The terms in the first two lines of Eq.\  \eqref{eq:covariance}
represent  the diagonal part of the covariance matrix. These include Statistical uncertainties
in redshifts ($\sigma^2_{z}$), in SNe IA magnitudes (owing to intrinsic variation ($\sigma^2_{\rm int}$)
and  gravitational lensing ($\sigma^2_{\rm lensing}$), in $m_B(\sigma^2_{m_B})$, $X_1 (\sigma^2_{X_1})$ and Color ($\sigma^2_{C}$)
and covariances between them ($\Sigma_{m_B,X_1}, \Sigma_{m_B,C}, \Sigma_{X_1,C}$) in each bin.
The terms in last line of  Eq.\  \eqref{eq:covariance} involving matrices
($V_0, V_a, V_b, V_{0a}, V_{0b}, V_{ab}$) correspond to the off-diagonal part
of the covariance matrix originating from statistical and systematic uncertainties. All these matrices are
given by JLA group and are extensively 
discussed in \cite{Betoule:2014frx,Wang:2016bba}.\\

We note from Eqs.\  \eqref{eq:muth}, \eqref{eq:dlrz}  and \eqref{eq:ez} that
evaluation of $\mu_{\rm th}^{(i)}$ requires values of parameters
$\Omega_r^{(0)}$, $\Omega_m^{(0)}$ and $\Omega_{de}^{(0)}$ and knowledge of
functional form of   equation of state (EOS) $w(z)$ of dark energy.
Since in a spatially flat universe $\Omega_r^{(0)} + \Omega_m^{(0)} + \Omega_{de}^{(0)} = 1$,  
neglecting the value of fractional density contribution of radiation at present epoch 
with respect to those from other components, we have $\Omega_{de}^{(0)} \approx 1 - \Omega_m^{(0)}$.
We may also choose different functional form of dark energy EOS which we 
denote by a general symbol $w(z;w_a,w_b,\cdots)$, where $w_a, w_b, \cdots$ denote the parameters
of the chosen functional dependence. On the other hand, the nuisance parameters $\alpha$ and $\beta$ 
enter in the expression for $\mu_{\rm obs}^{(i)0}$ (Eq.\  \eqref{eq:muobs}) 
and the covariance matrix as well (Eq.\ \eqref{eq:covariance}). 
The $\chi^2_{\rm SN}$ function, in Eq.\  \eqref{eq:chisqsn},
are minimised with respect to the  parameters  
$w_a$, $w_b$, $\Omega_m^0$, $\alpha$, $\beta$. 
The (best-fit) values of these parameters corresponding to the  minimum value of 
$\chi^2$ for different chosen models of dark energy EOS
are presented in Sec \ref{sec:results1}.\\

Besides SNe Ia data, compilation of measurements
of differential ages of the galaxies in GDDS, SPICES and VDSS surveys
gives measured values of  Hubble parameter at 15 different redshift values
\cite{ref:Riess,ref:simon1,ref:cabre1,ref:stern1}
The $\chi^2$ function for the analysis of this observational
Hubble data (OHD) may be defined as
\begin{eqnarray}
\chi^2_{\rm OHD}  &=& \sum_{i=1}^{15} {\Bigg{[}\frac{H(w_a,w_b,\Omega_m^{(0)};z_i)-H_{\rm obs}(z_i)}{\Sigma_i}\Bigg{]}}^2
\label{eq:chisqohd}
\end{eqnarray}
where $H_{\rm obs}(z_i)$ is the Observed value of the Hubble parameter at 
redshift $z_i$ with 1$\sigma$ uncertainty $\Sigma_i$ and $H(w_a,w_b,\Omega_m^{(0)};z_i)$
is its theoretical value evaluated by multiplying $E(z)$ in Eq.\  \eqref{eq:ez} by $H_0$.
Also, observation of Baryon Acoustic Oscillations (BAO) 
in Slogan Digital Sky Survey (SDSS)  provide measurement of
correlation function of the large sample of luminous red
galaxies. Using the detected acoustic peak value of a dimensionless standard ruler 
$A(z_1)$ corresponding to a  typical redshift $z_1 = 0.35$ may be determined. The theoretical expression
for the quantity $A(z_1)$ is given by
\begin{eqnarray}
A(w_a,w_b,\Omega_m^{(0)};z_1) &=&  \frac   {  \sqrt{  \Omega_m^{(0)}  }  }    {E^{1/3}(z_1)}
\Big{[} \frac{1}{z_1} \int_0^{z_1 }\frac{dz}{E(z)}\Big{]}^{2/3}
\label{eq:baorular}
\end{eqnarray}
where the parameters $w_a,w_b,\Omega_m^{(0)}$ enter in the above expression
though the function $E(z)$ (Eq.\  \eqref{eq:ez}). The observed value of the
standard ruler   $A_{\rm obs} \pm \Delta A$ is $0.469 \pm 0.017$ and the $\chi^2$-function for
the BAO data is taken as
\begin{eqnarray}
\chi^2_{\rm BAO} &=& \frac{\Big{[} A(w_a,w_b,\Omega_m^{(0)};z_1) -A_{\rm obs} \Big{]}^2}{(\Delta A)^2}
\label{eq:chisqbao}
\end{eqnarray}
To illustrate the impact of the Observational Hubble data and BAO data we have
performed a combined analysis of SNe IA, OHD and BAO data by 
minimising the total $\chi^2$ function
\begin{eqnarray}
\chi^2 & \equiv &\chi^2_{\rm SN} + \chi^2_{\rm OHD} + \chi^2_{\rm BAO}
\label{eq:totchisq}
\end{eqnarray}
with respect to the parameter set ($w_a$, $w_b$, $\Omega_m^0$, $\alpha$, $\beta$).
Results of the analysis are presented in Sec. \ref{sec:results1}.\\

\section{$k-$essence and Varying Dark energy}
\label{sec:kessence}
\begin{figure}[!t]
\begin{center}
\includegraphics[height=9cm, width=10cm]{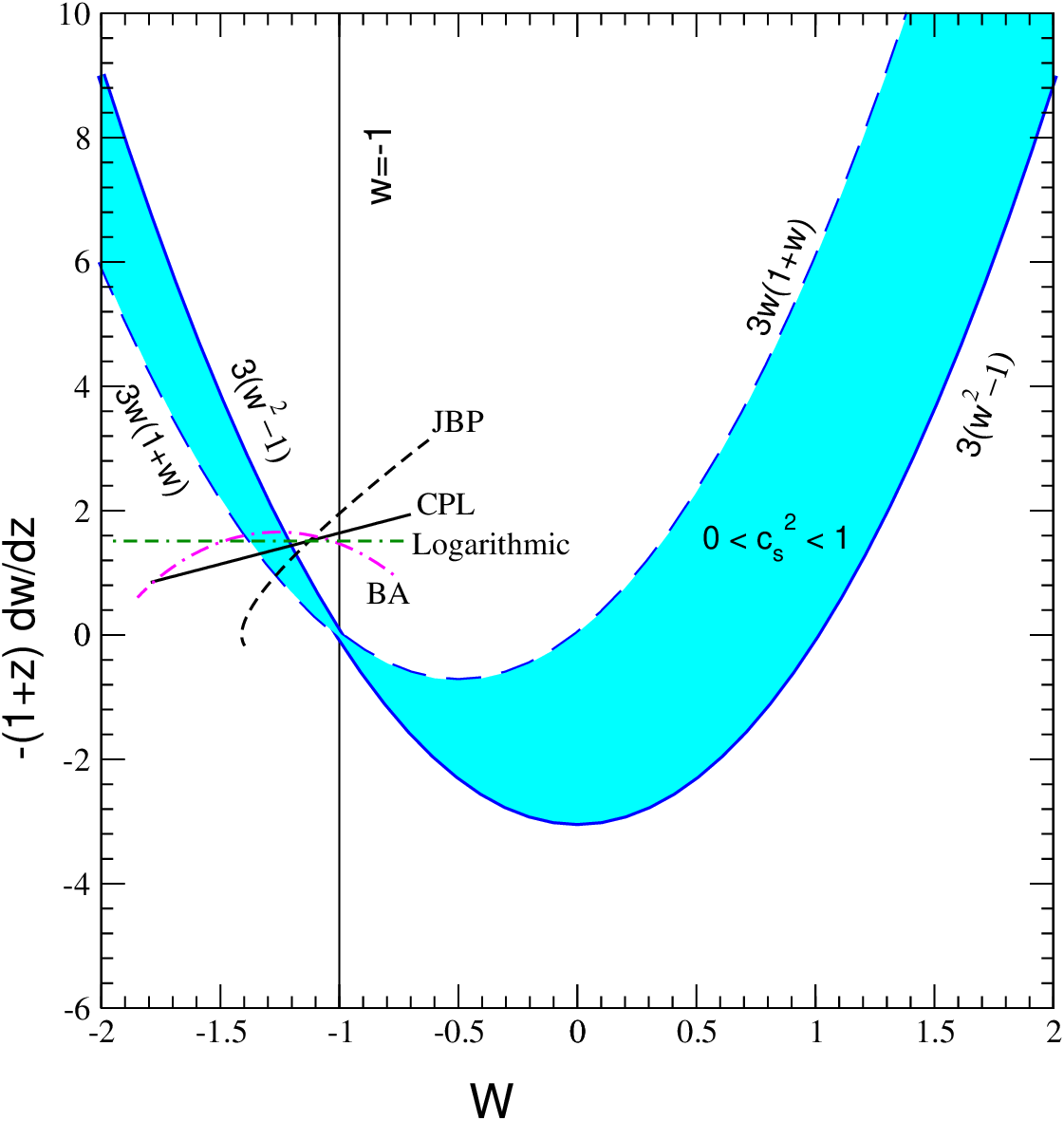}
\end{center}
\caption{\label{fig:1} Pictorial representations of the conditions for 
$c_s^2>0$ and $c_s^2<1$  in the 
parameter space spanned by  $w$ and $\left(-(1+z)dw/dz\right)$
(See Eqs.\ (\ref{eq:cond1}) and (\ref{eq:cond2})).
For $w<-1$, the   region lying  above the dotted line ($3w(1+w)$)
 and for $w>-1$ region lying  below
 the dotted line corresponds to $c_s^2>0$. The shaded region corresponds to
 the bound $0<c_s^2<1$. The curves at the best-fit ($w_a,w_b$) points
 corresponding to
 different parametrisations of $w(z)$  in different varying dark energy models
 are also shown.
}
\end{figure}

In this work we investigate realisation of dark energy with varying equation of state in terms
of $k-$essence scalar field models. We assume dark energy  represented in terms of a  
homogeneous scalar field
whose dynamics is driven by a $k$-essence Lagrangian with
constant potential. In this context, we give below a brief outline of 
basic equations of  $k-$essence model \textit{i.e.}
\begin{eqnarray}
L  &=& V(\phi)F(X) =  p \label{eq:s2}\\
\rho  &=& V(\phi)(2XF_X - F) \label{eq:s3}
\end{eqnarray}
where $L$ is the $k$-essence Lagrangian, $\rho$ and $p$ 
respectively represent energy density and pressure of dark energy. 
$F(X)$ is  a function of $X$, where
$X =  \frac{1}{2}\nabla_{\mu}\phi\nabla^{\mu}\phi$, $F_X \equiv dF/dX$ and
$V(\phi)$ represents the potential.
In a flat FLRW spacetime background, $\rho$ and $p$ are related 
by the continuity equation
\begin{eqnarray}
\dot{\rho} + 3H(\rho+p) &=& 0\,,
\label{eq:conti}
\end{eqnarray}
where $H = \dot{a}/a$ is the Hubble constant and $a(t)$ is the scale factor.
For a homogeneous  scalar field $\phi$, in a flat FLRW spacetime
background, we have $X = \frac{1}{2}\dot{\phi}^2$.
 We consider $k-$essence models with constant potential
 $V(\phi)=V$ which ensures existence of scaling relation
\cite{scale1,scale2}
\begin{eqnarray}
XF_X^2 &=& C a^{-6} 
 \label{eq:s4}
\end{eqnarray}
where $C$ is a constant. 
 The  equation of state of dark energy represented by $k-$essence 
field is given by
\begin{eqnarray}
w &=& \frac{p}{\rho} = \frac{F}{2XF_X - F}
\end{eqnarray}

 The issue of causality in the context of $k-$essence scalar field theories with
Lorentz invariant action of the form $S = \int d^4x \sqrt{-g} L(\phi,X)$ where $g$ is the 
determinant of the FRW metric considered here and $L$ is the Lagrangian in Eq.\  (\ref{eq:s2}) has been
discussed in detail in \cite{vik1,vik2}. Variation of the action with respect to the scalar field 
gives the equation of motion of the scalar field $\phi$ as 
\begin{eqnarray}
G^{\mu\nu}\nabla_\mu\nabla_\nu \phi + 2X\frac{\partial}{\partial X} \left(\frac{\partial L}{\partial \phi}\right)
 - \frac{\partial L}{\partial \phi} &=& 0
\label{eq:phieom}
\end{eqnarray}
where, the effective metric
$G^{\mu\nu} \equiv L_{X}g^{\mu\nu} + L_{XX} \nabla^\mu\phi \nabla^\nu\phi$ with  $L_X$ and $L_{XX}$
denoting $\partial L / \partial X$ and  $\partial^2 L / \partial X^2$ respectively, has a Lorentzian 
structure and describes the time evolution of the system if the following condition is satisfied \cite{vik1,newref1,newref2,newref3}
\begin{eqnarray}
1 + 2X \frac{L_{XX}}{L_X} > 0
\label{eq:cond6}
\end{eqnarray}
Now  introducing   $c_s^2$ as 
\begin{eqnarray}
c_s^2 &\equiv & \left[1 + 2X \frac{L_{XX}}{L_X}\right]^{-1}
\label{eq:cssqdef}
\end{eqnarray}
it has been shown in \cite{newref4}, that for $X (=\frac{1}{2}\dot{\phi}^2, \rm{in~our~case})>0$, 
the quantity $c_s^2$ plays the role of sound speed squared for propagation of small perturbations.
However, in the context of spatially flat FRW universe, with small perturbations, neglecting vector perturbations
which decay as $a^{-2}$, the metric may be
written as 
\begin{eqnarray}
ds^2 &=& (1 + 2\Phi) dt^2 - a^2(t) [(1-2\Phi)\delta_{ij} + h_{ij}]dx^i dx^j
\label{eq:per1}
\end{eqnarray}
where $\Phi$ (the gravitational Newtonian potential) is the scalar perturbation
and $h_{ij}$ is traceless transverse perturbations. From the standard results of cosmological 
perturbation theory  \cite{newref4,mukha1,mukha2}, it follows that 
perturbations in the k-essence field $\delta\phi$,
which are gauge invariant are connected with the scalar metric perturbations
and the dynamics of cosmological perturbation may described by the action of the 
form
\begin{eqnarray}
S_c &=& \frac{1}{2} \int d^3x d\eta \left[ \left(\frac{dv}{d\eta}\right)^2 - c_s^2 (\nabla v)^2 - m_c^2 v^2\right]
\label{eq:per2}
\end{eqnarray}
where $\eta\equiv\int dt/a(t)$ is the conformal time, 
$v \equiv \sqrt{\frac{d\rho}{dX}}a (\delta\phi + \frac{1}{{\cal H}} \frac{d\phi}{d\eta}\Phi)$, ${\cal H} = (1/a)(da/d\eta)$,
$m_c^2 \equiv -(1/z)(d^2z/d\eta^2)$ with $z \equiv \sqrt{\frac{d\rho}{dX}}\frac{a}{\cal H}\frac{d\phi}{{d\eta}}$ and 
the quantity $c_s^2$ representing sound speed squared for propagation of small perturbations in Eq.\  (\ref{eq:per2})
is given by 
\begin{eqnarray}
c_s^2 &=& \frac{dp/dX}{d\rho/dX}  
\label{eq:per3}
\end{eqnarray}
A derivation for the above formula from an effective hydrodynamical description of the system
is also obtained in \cite{vik2}.

Therefore, for the classical solutions $F(X)$ of the scaling relation Eq.\  \eqref{eq:s4} to be
stable against small perturbations of the background energy density,
the square of adiabatic sound speed   should be positive ($c_s^2>0$).
On the other hand, causality arguments require that this speed of propagation small 
perturbations of the background  should not exceed the
speed of light, implying $c_s^2<1$ \cite{dePutter,Abramo,Cardenas}.  
In the context of $k-$essence model, using Eqs.\  (\ref{eq:s2}) and  (\ref{eq:s3})
in  (\ref{eq:per3}) we have
\begin{eqnarray}
c_s^2 &=&   \frac{F_X}{2XF_{XX} + F_X }
\label{eq:s9}
\end{eqnarray}
Using $1 /a = 1 + z$ (where $z$ is the redshift and value of scale factor 
$a$ at present epoch is normalised to unity) in $H = \dot{a}/a$ we have $dt = -dz/(1+z)H$.
Exploiting this result and transforming time dependences 
of $\rho$, $p$ and $w$ to their $z-$dependences in  Eq.\  \eqref{eq:conti}
we may also express the sound speed squared as a function of 
redshift $z$ as
\begin{eqnarray}
c_s^2 & = &   \frac{3w(1+w) +(1+z) dw/dz}{3(1+w)}
\label{eq:s10}
\end{eqnarray}
We note from Eq.\  \eqref{eq:s10}   that the bound $c_s^2>0$ corresponds to
\begin{eqnarray}
 \mbox{either}\quad &&-(1+z) dw/dz  < 3w(1+w), w  >-1  \nonumber \\
 \mbox{or}\quad &&-(1+z) dw/dz > 3w(1+w) , w  <-1
\label{eq:cond1}
\end{eqnarray}
and the bound $c_s^2<1$ corresponds to
\begin{eqnarray}
\mbox{either}\quad && -(1+z) dw/dz  > 3(w^2-1), w  >-1   \nonumber \\
\mbox{or}\quad && -(1+z) dw/dz  < 3(w^2-1) , w  <-1
\label{eq:cond2}
\end{eqnarray}
Whether the physical bound $c_s^2>0$ (or $0<c_s^2<1$)  
is realised for any chosen functional form of 
equation of state $w(z)$ may be verified by 
checking the   conditions given in
Eq.\  \eqref{eq:cond1} (or Eqs.\  \eqref{eq:cond1}and  \eqref{eq:cond2}).
These conditions 
are pictorially demonstrated 
 in Fig.\ \ref{fig:1} (see \cite{dePutter} for details) where we have studied 
 effect of the conditions on
  the plane  spanned by quantities
 $w$ and $\left(-(1+z)dw/dz\right)$.
The shaded region marked in the figure is bounded by two curves
$\left(-(1+z)dw/dz\right)=3w(1+w)$ (dashed line) and 
$\left(-(1+z)dw/dz\right)= 3(w^2-1)$ (solid line). 
We have also shown the $w=-1$ line in the plane.
From Eq.\  \eqref{eq:cond1} and \eqref{eq:cond2} 
we see that $c_s^2>0$ corresponds to
the region in the plane which is below the dashed line for $w>-1$ and above the dashed line for $w<-1$.
The condition $c_s^2<1$, on the other hand, corresponds to
the region  lying  above the solid line for $w>-1$ and below the solid line for $w<-1$. 
The shaded region, bounded between these two lines in the plane, therefore corresponds 
$0<c_s^2<1$ for the entire range of  values of $z$ accessible in the 
observations considered here.\\

\begin{table}[t!]
\begin{center}
 \begin{tabular}{l|l|l}
\hline
Model & $w(w_a,w_b;z)$ & $Y(z) \equiv \exp\left[3\int_0^z dz^\prime \frac{1+w(z^\prime)}{1+z^\prime}\right]$ \\
&&\\
\hline
CPL \cite{CPL1} & $ w_a + w_b \frac{z}{1+z}$ & $(1 + z)^{3(1 + w_a + w_b)}\exp{\Big{(}\frac{-3w_az}{1+z}\Big{)}}$\\
\hline
JBP \cite{JBP1},\cite{JBP2} & $ w_a + \frac{w_bz}{(1+z)^2}$& $(1 + z)^{3(1 + w_a)}\exp{\Big{(}\frac{3w_bz^2}{2(1+z)^2}\Big{)}}$\\
\hline
BA \cite{Barboza1},\cite{Barboza2} & $ w_a + \Big{(}\frac{w_bz(1+z)}{1+z^2}\Big{)}$ & $ (1 + z)^{3(1 + w_a)}(1 +z^2)^{\frac{3w_b}{2}}$\\
\hline
Logarithmic \cite{Sangwan} & $ w_a + w_b\log(1+z)$ & $(1 + z)^{3(1 + w_a + \frac{w_b}{2}\log(1+z))}$\\
\hline
\end{tabular}
\end{center}
\caption{\label{tab:1} 
Functional forms of equation
of state $w(z)$   of dark energy
 as used in different  varying dark energy Models.
Expressions for corresponding $z$-dependences of 
dark energy density, expressed through the function $Y(z)$ (Eq.\  (\eqref{eq:ss1})) are also given.}
\end{table}

Variation in the dark energy density $\rho(z)$   may be expressed
in terms of variation of  dark energy equation of state  $w(z)$.
Expression for a general functional form of $w(z)$, in principle, involves
infinite number of parameters. 
However, for a practical analysis its   effective 
to express the functional dependence  $w(z)$, in terms a small number of parameters and consider 
different forms of  parametrizations of $w(z)$ . We  consider  here 4 different models,
often used in literature in the context of varying dark energy, {\it viz.} CPL\cite{CPL1}  , JBP\cite{JBP1},\cite{JBP2}, BA(\cite{Barboza1},\cite{Barboza2})
and Logarithmic model \cite{Sangwan}.
Each of the models  uses   
a specific functional form of $w(z)$ expressed in terms of two  parameters ($w_a$ and $w_b$)
and are listed in Table\ \ref{tab:1} where we have
also given   functional form of the quantity
$ Y(z) \equiv \exp\left[3\int_0^z dz^\prime \frac{1+w(z^\prime)}{1+z^\prime}\right]$ which 
gives the $z-$dependence of the corresponding dark energy density $\rho(z)$ (see Eq.\  \eqref{eq:ss1}).
In the plane of  Fig.\ \ref{fig:1}, we have also shown the curves
representing the different parametrisations of $w(z)$ 
for the best-fit values of parameters $w_a$ and $w_b$ obtained from the analysis of
observational data (see Sec.\ \ref{sec:results1}).\\

We finally exploit the equations of $k-$essence models to reconstruct the
form the function $F(X)$.
 Using Eq.\  \eqref{eq:conti} we    express energy density as a function of redshift as
\begin{eqnarray}
\rho &=& \rho^{(0)} Y(z)\,,\quad \mbox{where }Y(z)= \exp\left[3\int_0^z dz^\prime \frac{1+w(z^\prime)}{1+z^\prime}\right]  
\label{eq:ss1}
\end{eqnarray}
where $\rho^{(0)}$ corresponds to value of dark energy density at present epoch $(z=0)$.
Using Eqs.\  \eqref{eq:s2}, \eqref{eq:s3}, \eqref{eq:s4} and \eqref{eq:ss1} we obtain,
\begin{eqnarray}
\left(\frac{4CV^2}{{\rho^{(0)}}^2} \right)X &=& \frac{  Y^2(z) (1 + w(z))^2}{  (1+z)^6}
\label{eq:ss2}
\end{eqnarray}
Writing $p=\rho w$ in  Eq.\ \eqref{eq:s2} and then using Eq.\ \eqref{eq:ss1} we have,
\begin{eqnarray}
\left(\frac{V}{\rho^{(0)}} \right)F(X) &=&     Y(z)w(z)  
\label{eq:ss3}
\end{eqnarray}
For a given form of the equation of state $w(z)$, the right hand sides
of Eqs.\  \eqref{eq:ss2} and \eqref{eq:ss3} may be evaluated numerically at
each $z$. 
Eliminating $z$ from both the equations one may obtain the  
$X$-dependence of the function $F(X)$ corresponding to a given form of $w(z)$.
The  dependences of $F(X)$ on $X$ obtained from the analysis of observational data
are shown and described in Sec.\ \ref{sec:results1}.

\section{Results of analysis of observed data}
\label{sec:results1}
\begin{figure}[t]
\begin{center}
\includegraphics[width=14cm, height=9cm]{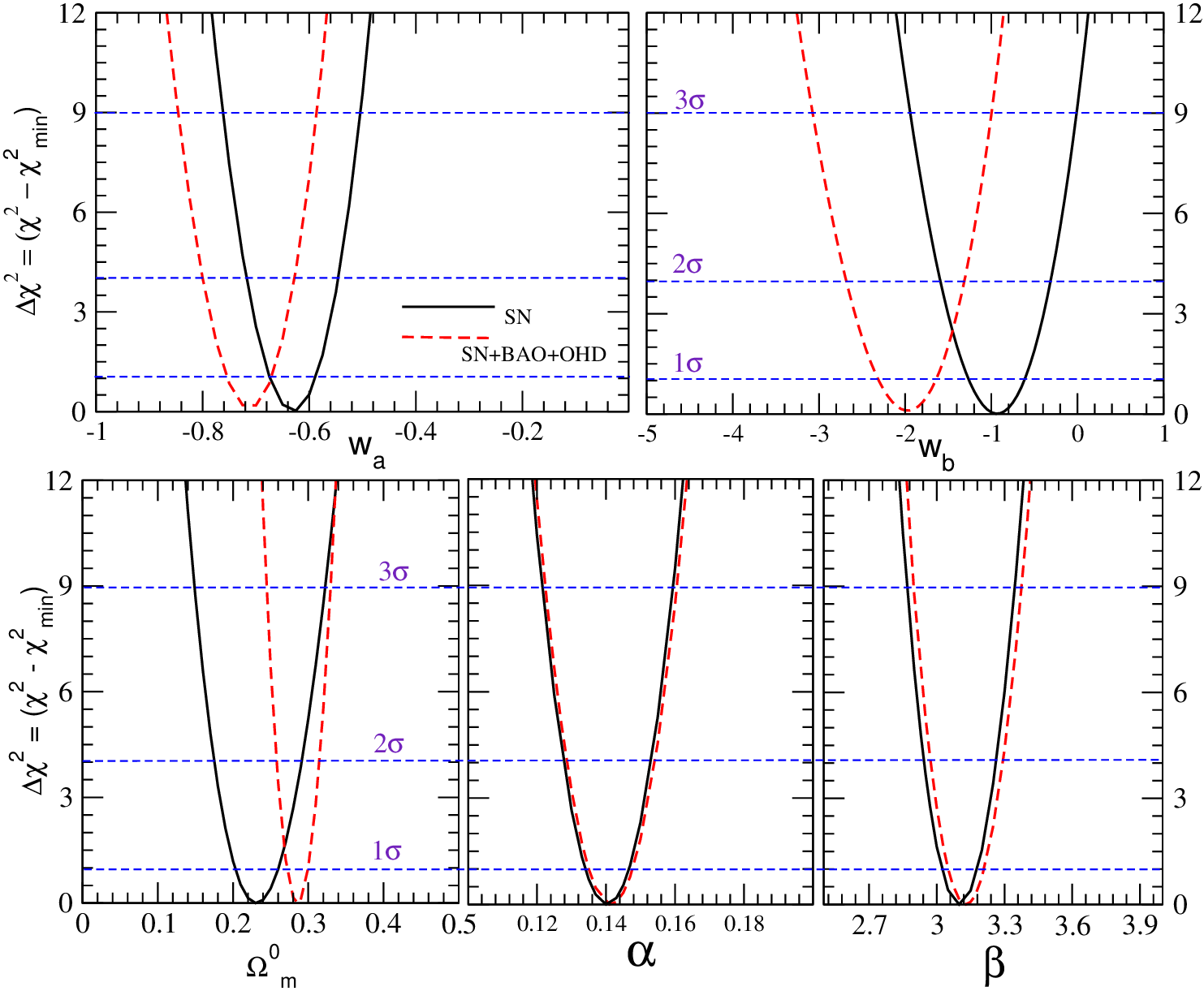}
 \end{center}
\caption{\label{fig:2a} 
Plots of $\chi^2 - \chi^2_{\rm min}$ as a function
of each individual parameters of the set ($w_a$, $w_b$, $\Omega_m^0$, $\alpha$ and $\beta$). The CPL parametrisation of $w(z)$ has been used. In each of
the plots, depicting variation of $\chi^2$ with one of the parameters at a time,
values of the other parameters are kept fixed at their respective best-fit values
as given in Tab.\ \ref{tab:2} (solid lines for best-fits of SNe Ia data alone and dotted lines for the best-fits from the combined analysis of SNe Ia, BAO and OHD).
The values of $\Delta \chi^2$ \textit{viz.} 1, 4 and 9,  
corresponding respectively to  one parameter confidence levels 
of 1$\sigma$, 2$\sigma$
and 3$\sigma$ are shown by dotted horizontal lines.}
\end{figure} 
\begin{figure}[t]
\begin{center}
\includegraphics[width=12cm, height=12cm]{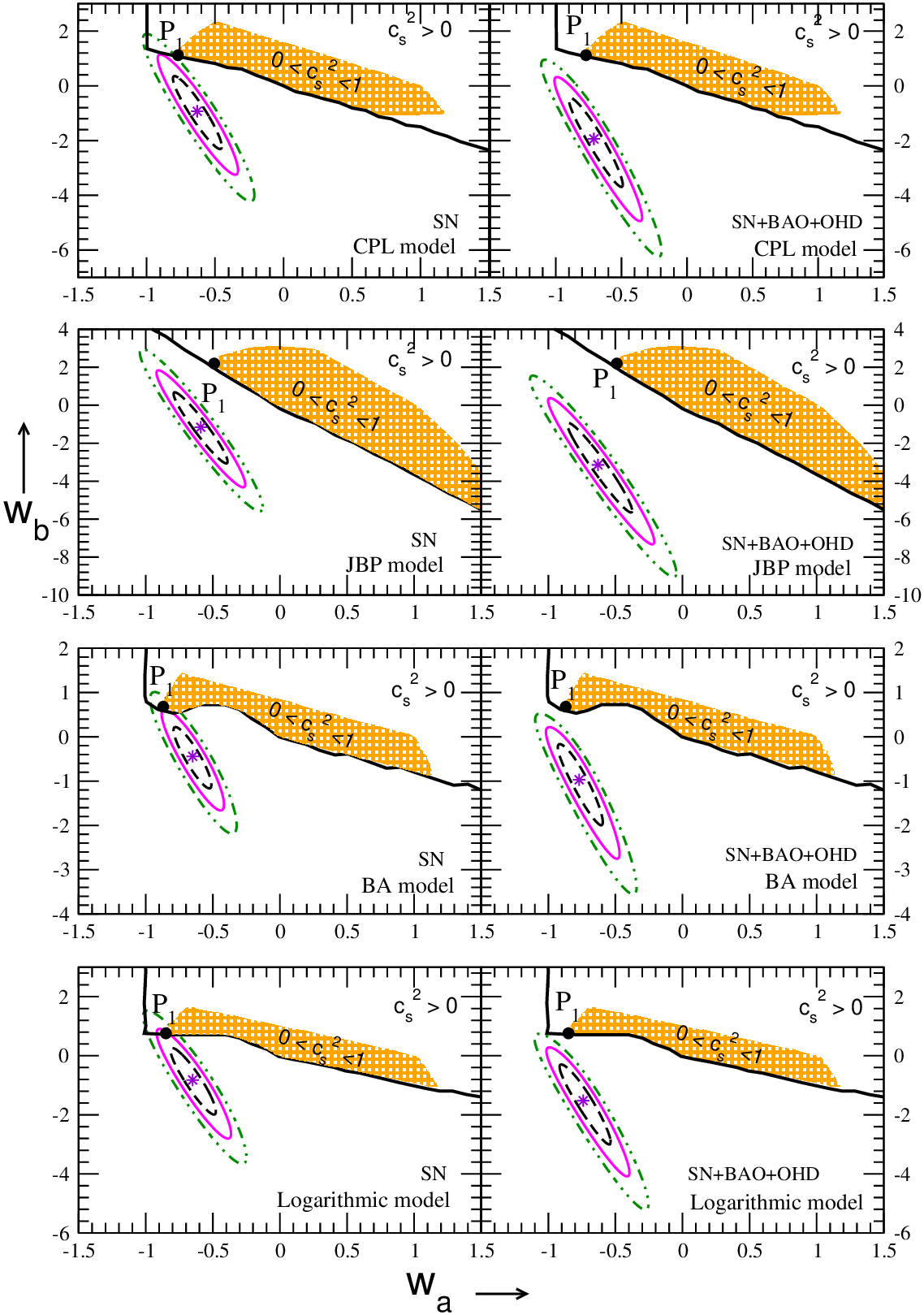}
 \end{center}
\caption{\label{fig:2}
Regions of   $w_a - w_b$ parameter space 
allowed at  $1\sigma$, $2\sigma$ and $3\sigma$  confidence limits 
from analysis of SNe Ia data (left panel) and SNe Ia + BAO + OHD (right panel).
The results for 4 different types of parametrisations of $w(z)$ are shown in
different rows. The corresponding best-fit point and the point $P_1$ are
also shown (see text for details). The region above the black solid line
corresponds to $c_s^2>0$. The shaded region 
in the figure corresponds to the bound $0<c_s^2<1$.}
\end{figure}

\begin{figure}[!t]
\begin{center}
\includegraphics[scale=.5]{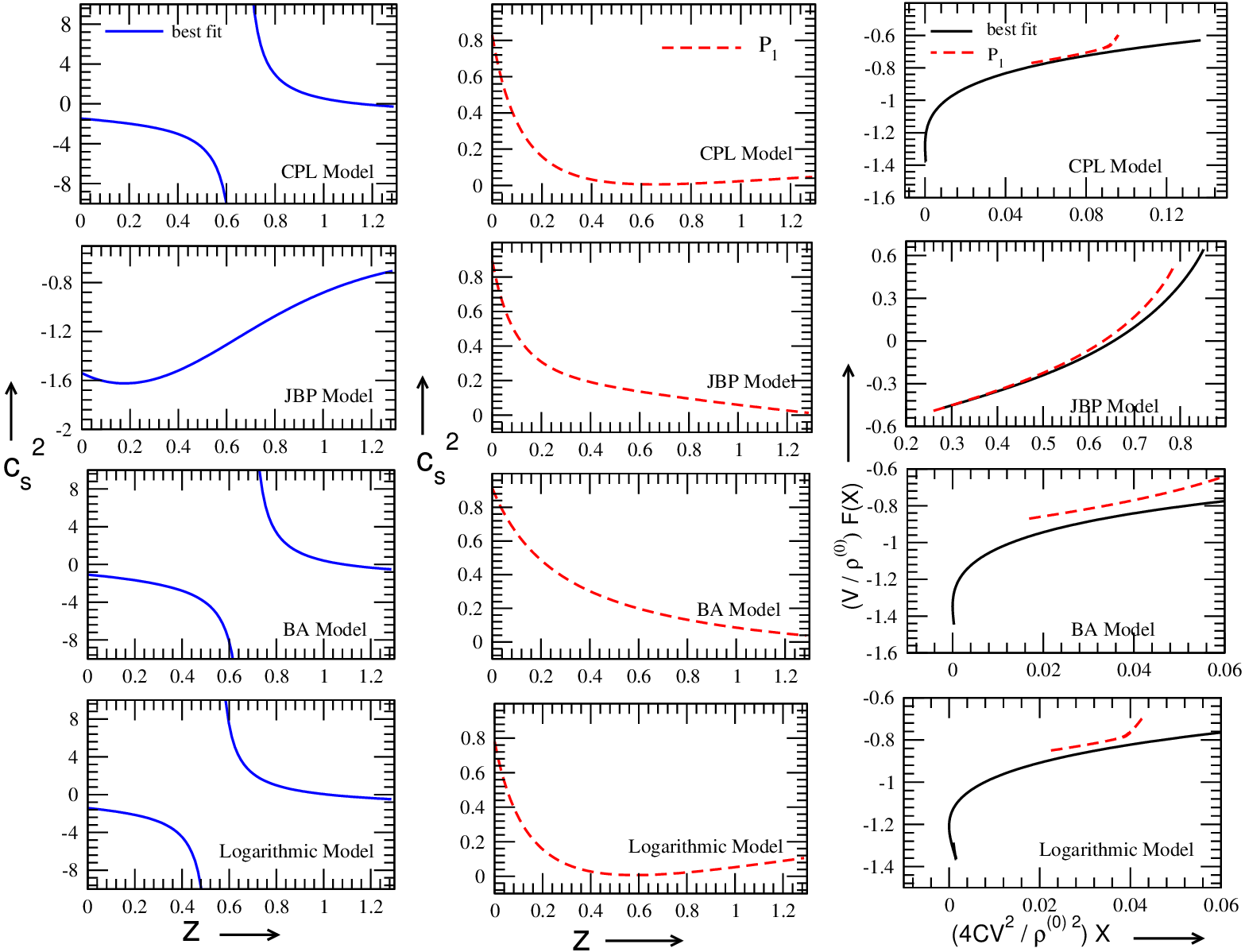}
\end{center}
\caption{\label{fig:3}
Plot of $c_s^2$ vs $z$ for different parametrisations of $w(z)$ 
at best-fit (right panel) from SNe Ia data and 
at $P_1$ (middle panel) (see text for details). 
Corresponding plots of $F(X)$ vs $X$ reconstructed at best-fit and $P_1$
are shown in right panel.}
\end{figure}
\begin{table}[!t]
\begin{center}
\begin{tabular}{c|ccccccccc}
Data &&&&&&&& Range of $z$  \\
 Set &Model & $w_a$ & $w_b$ & $\Omega_m^0$ & $\alpha$ & $\beta$ & $\chi^2$/DOF & for which  \\
&&&&&&&&   $0<c_s^2<1$ & $P_1 (w_a,w_b)$ \\
 \hline
&CPL & -0.63  & -0.93  & 0.23 & 0.14 & 3.1 & 685.42/735 & 0.93 - 1.15 & (-0.77,1.15)\\
&JBP & -0.59  & -1.16  & 0.20  & 0.14 & 3.1 & 685.39/735 & - &(-0.49,2.22)\\
SNe Ia &BA & -0.65  & -0.44 & 0.22 & 0.14  & 3.1 & 685.48/735 & 0.94 - 1.10&(-0.87,0.65)\\
&Log. & -0.65 & -0.82 & 0.24  & 0.14  & 3.1 & 685.44/735 & 0.79 - 1.01&(-0.85,0.76)\\
 \hline
 \hline
&CPL & -0.71 & -1.94 & 0.29 & 0.14 & 3.13 & 708.26/751 & $0.35 - 0.47$ &\\
SNe Ia &JBP & -0.63  &  -3.14 & 0.28 & 0.14 & 3.13 & 707.65/751 & $0.31 - 0.39$ &\\
+ BAO   &BA & -0.77  & -0.97 & 0.29 & 0.14 & 3.13 & 708.86/751 & $0.45 - 0.53$& \\
+ OHD &Log.& -0.74 & -1.52 & 0.29 & 0.14 & 3.13 & 708.56/751 & $0.38 - 0.51$&\\
 \hline
 \end{tabular}
\end{center}
\caption{\label{tab:2}  Best-fit values of parameters for different models from analysis of 
SNe Ia data alone and SNe + BAO + OHD. The values of $\chi^2$ ($\chi^2_{\rm min}$) at the best-fit point 
per DOF (degrees of freedom) are also shown. The range of $z$ for which the value
of $c_s^2$ evaluated at best-fit lies between 0 and 1 are also shown.
Last column shows the values of ($w_a,w_b$) corresponding to point $P_1$ (see text for details).}
\end{table}

\begin{table}[!t] 
\begin{center}
\begin{tabular}{|c|c|l|l|l|l|l|}
\hline
 Data   & Model & 1$\sigma$ \& 3$\sigma$   & 1$\sigma$ \& 3$\sigma$  &   1$\sigma$ \& 3$\sigma$ &  1$\sigma$ \& 3$\sigma$  & 1$\sigma$ \& 3$\sigma$   \\
 Set         &   &   range of $w_a$          & range of $w_b$ & range of $\Omega_m^0$  & range of $\alpha$  & range of $\beta$ \\
 \hline
           & CPL & [-0.67,-0.58]        & [-1.25,-0.60]   & [0.20 ,0.26 ]                        &                    &  \\
           &   &  \& [-0.76,-0.50]      &  \&[-1.95,0.01]  &                       \& [0.15 ,0.32 ] &                    &  \\
           \cline{2-5}
SNe Ia     & JBP & [-0.63,-0.55]   & [-1.51-0.77]  &    [0.17 ,0.23 ]                    &                    &  \\
           &   & \& [-0.70,-0.47]    & \& [-2.27,-0.05]  & \&                       [0.12,0.30]  &                    &  \\
           \cline{2-5}        
           & BA  & [-0.69,-0.61]     & [-0.64-0.23]  &    [0.19 ,0.25 ]                      &  [0.13,0.15]                  & [0.24,0.25]  \\
           &    &  \&[-0.77,-0.52]     & \&[-1.10,0.17] &  \&                                             [0.12,0.30]  &  \&   [0.12,0.16]                                                           & \& [0.23,0.26]  \\  
                \cline{2-5}         
           & Log.&  [-0.69,-0.60]      &  [-1.12-0.48]&   [0.22,0.27]                   & 
            &  \\
          &     &  \& [-0.78,-0.51]      &  \& [-1.71,0.05]  & \&                                                      [0.16,0.33]    &         &  \\
\hline\hline
            & CPL & [-0.75,-0.67]        & [-2.3,-1.63]           &                        [0.27 ,0.30] &                    &  \\
           &   &  \& [-0.84,-0.58]      &  \&[-3.07,-0.98] &                       \& [0.24 ,0.33]   &                    &  \\
           \cline{2-5}
SNe Ia     & JBP & [-0.65,-0.58]          & [-3.39,-2.6] &                                       [0.27,0.30]  &                    &  \\
+BAO         &   & \& [-0.74,-0.49]      & \& [-4.25,-1.82]  &                       \&[0.24 ,0.33]   &                    &  \\
          \cline{2-5}        
 +OHD           & BA  & [-0.82,-0.74]      & [-1.24,-0.79 ]&                                [0.27 ,0.30 ] & [0.13 ,0.15]                   & [0.24,0.25]  \\
           &    &  \&[-0.91,-0.65]        & \&[-1.80,-0.32]   &                       \&[0.24 ,0.33 ]   & \&  [0.12 ,0.16]                    & \& [0.23,0.26] \\  
                \cline{2-5}         
           & Log.&  [-0.78,-0.70]        &  [-1.84,-1.22] &                   [0.27 ,0.30 ] &  & \\
          &     &  \& [-0.87,-0.61]      &  \& [-2.53,-0.68]                      & \& [0.24 ,0.33]        &           &  \\
\hline
 \end{tabular}
\end{center}  
\caption{\label{tab:2a} 1$\sigma$ and 3$\sigma$ confidence limits of the
each of the individual parameters of the set ($w_a$, $w_b$, $\Omega_m^0$, $\alpha$ and $\beta$) for different parametrisations of $w(z)$ (See text for details.) }
\end{table}

In this section we present the results of 
analysis of the observational data using the 
methodology described in Sec.\ \ref{sec:analysis}.
We investigate  implications of the observations in the context
of the varying dark energy models listed in Table\ \ref{tab:1}.
For each of the models we obtain the best-fit values of the parameters 
($w_a$, $w_b$) along with their
allowed domains at different confidence limits from the analysis.  
We perform  analysis of SNe Ia data alone and also 
a combined analysis of data
from SNe Ia, BAO and OHD (discussed in Sec.\ \ref{sec:analysis}) where 
we freely vary the parameters $w_a$, $w_b$, $\Omega_m^0$, $\alpha$ and $
\beta$ to find their best-fit values (corresponding to minimum
value of  $\chi^2$ in Eq.\  (\ref{eq:totchisq})). The obtained best-fit values of the above parameters
for different models are presented in Table\ \ref{tab:2}. \\

We also find the ranges of the individual parameters allowed
at different confidence levels  from the analysis of the observational data. To obtain this,
we find the variation of $\chi^2$ with each of the parameters of the set
\{$w_a$, $w_b$, $\Omega_m^0$, $\alpha$, $\beta$\} at a time, keeping values of
all other parameters fixed at their respective best-fit values.
The confidence interval for the single parameter may then
be obtained from the distribution of the function $\Delta\chi^2 \equiv \chi^2 - \chi^2_{\rm min}$ \cite{nures}.
The range of values of the parameter for which   $\Delta\chi^2 \leqslant 1$, $\Delta\chi^2 \leqslant 4$ and $\Delta\chi^2 \leqslant 9$ 
respectively correspond to  
$1\sigma$ (68.3\% Confidence Level (C.L)), $2\sigma$ (95.4\% C.L)
and $3\sigma$ (99.73\% C.L) \cite{nures} allowed intervals of the parameter.
We have shown in  Fig.\ \ref{fig:2a}
the nature of dependence of $\Delta\chi^2 \equiv \chi^2 - \chi^2_{\rm min}$ 
on each of the individual parameters. For demonstrative purpose, we have shown the plot for CPL model only. 
However, the  obtained 1$\sigma$ and $3\sigma$ ranges  of the individual parameters for different models
of parametrisations of $w(z)$, are given in Tab.\ \ref{tab:2a}.\\

For a comparative study of the different parametrisations of $w(z)$,
we have also displayed the joint confidence region in the parameter space of 
$w_a$ and $w_b$.  To obtain this we keep the other parameters ($\omega_m^0$, $\alpha$, $\beta$)
at their respective best-fit values and obtain domains in 
$w_a - w_b$ parameter space for which evaluated values of $\chi^2$  
lie in the domain $\chi^2 = \chi^2_{\rm min} + \Delta\chi^2$.  The 1$\sigma$,
2$\sigma$ and 3$\sigma$ joint confidence region of  two parameters ($w_a$ and $w_b$)
correspond to $\Delta \chi^2 <2.30$,  $\Delta \chi^2 <6.17$ and $\Delta \chi^2 <11.8$ respectively \cite{nures}.
The obtained joint confidence regions in $w_a - w_b$ parameter space for the different $w(z)$ parametrisations are shown in   
Fig.\ \ref{fig:2}. The corresponding best-fit   points are also shown in the parameter space.\\

In the context of $k-$essence model of dark energy, we  investigate,
to what extent the condition  $0<c_s^2<1$  is favoured from observational data 
for different parametrisations of the dark energy equation of state $w(z)$.
We used the  different parametrisations of  $w(z)$ 
in Eqs.\   \eqref{eq:cond1} and  
\eqref{eq:cond2} to find the range of values of the parameters $w_a$ and $w_b$ for which 
the condition  $0<c_s^2<1$ is satisfied for all $z$ within
the domain of observations.
 For each of the  parametrisations of $w(z)$ mentioned 
in Table\ \ref{tab:1}, this range is shown by  a shaded region
in $w_a - w_b$ plane in Fig.\ \ref{fig:2}. 
The regions corresponding to $c_s^2>0$ only  are also shown for each model.
We then observe that
for all the models,
the  shaded region corresponding to $0<c_s^2<1$ 
has no overlap with the region  bounded by $1\sigma$ 
contour allowed from analysis of SNe Ia data alone.
For BA and Logarithmic models the overlap is seen when
one considers allowed ranges at and beyond  $\sim 2\sigma$ confidence limits.
This implies that the physical bound  $0<c_s^2<1$   for the entire range of values of $z$ probed by the 
observed data considered here, is favoured from observational data (SNe Ia only) at and above $\sim 2\sigma$ 
confidence level if we consider parametrisations of  $w(z)$ as in BA and Logarithmic models.
For CPL parametrisation the physical bound is  disfavoured below $\sim 3\sigma$
and with JBP its disfavoured  upto  even higher confidence limits. 
For all the different parametrisations of $w(z)$
 the physical bound is disfavoured to a larger 
 extent from a combined analysis of SNe Ia , BAO and OHD.
In the $w_a - w_b$ 
parameter space shown in Fig.\ \ref{fig:2}, we have also marked a point $P_1$  
in the shaded region corresponding to $0<c_s^2<1$, at which the value of $\chi^2$
is closest to the value of $\chi^2_{\rm min}$ for the corresponding model.
Thus $P_1$ refers to the maximally favoured values of parameters $w_a$ and $w_b$ from
observational data for which $c_s^2$ lies between 0 and 1 for all $z$ values. The values of 
($w_a,w_b$) corresponding to $P_1$ are shown in the last column of table\ \ref{tab:2}.\\

At the best-fit values of parameters $w_a$ and $w_b$
obtained from the combined analyses of observational data from Sne IA, BAO and OHD for different $w(z)$ parametrisations we
have shown the variation of sound speed squared $c_s^2(z)$ with redshift $z$ in left panel of
Fig.\ \ref{fig:3}. We see that at the best-fit values of the
parameters the calculated value of $c_s^2$ lies within its physical bound $0<c_s^2<1$ only
for a vary narrow range of values of $z$. These ranges are also shown in Table\ \ref{tab:2}.
The variation of $c_s^2$ at values of $w_a, w_b$ corresponding to the
point $P_1$ are shown for different models in the middle panel of Fig.\ \ref{fig:3}.
These correspond to a monotonous variation of $c_s^2$ with $z$ within its physical bound $0<c_s^2<1$
imposed by causality and stability. \\

In Sec.\ \ref{sec:kessence} we discussed the methodology to
reconstruct the form of function $F(X)$ for different 
form of parametrisations of $w(z)$. 
We have shown in right panel of Fig.\ (\ref{fig:3}) the obtained dependences of $F(X)$ on $X$, 
for each of the models  at the corresponding 
best-fit values of the parameters $(w_a,w_b)$.
The Figure shows that for JBP model $F(X)$ has a monotonous dependence of $X$
whereas for the other models (CPL, BA and Logarithmic) the function is
double valued in a certain domain of $X$.  \\

  As discussed earlier,  
the cosmological parameters that enter  into our analysis are $\Omega_m^0$ and parameters $w_a$ and $w_b$
which parametrise the equation of state of dark energy. We have obtained best-fit 
values of the parameters from analysis of the Sne Ia observational data. 
However, for a cosmographic analysis, 
which is  a model independent way 
of processing cosmological data, 
the chosen parameter set is different. 
Basic aspects of cosmographic methodology and results of  
cosmographic analysis of SNe data are discussed in \cite{cg1,cg2,cg3}. 
Here we briefly discuss the cosmography constructed from DE equation of state. 
We also qualitatively compare results of our analysis with
the features of results of cosmographic analysis in terms of cosmographic parameters.\\

Neglecting the contribution to the present day energy density due to radiation $\Omega_r^0$
in Eq.\  (\ref{eq:ez}) we can express the EOS parameter of dark energy as 
\begin{eqnarray}
\omega(z) &=& -1 + \frac{1}{3}\frac{\big{[}E(z)^2-\Omega_m^0(1+z)^3\big{]}^\prime(1+z)}{E(z)^2-\Omega_m^0(1+z)^3} 
\label{eq:wzcg}
\end{eqnarray}
where, $^\prime$ in above equation denotes derivative with respect to $z$. We consider two 
cosmographic parameters: the  deceleration parameter $q(z)$ and the jerk parameters $j(z)$
which are defined in a model independent way as 
\begin{eqnarray}
q(z) = -\frac{a\ddot{a}}{\dot{a}^2}\quad \mbox{and} \quad j(z) = \frac{\dddot{a}a^2}{\dot{a}^3}
\end{eqnarray}
These parameters are relevant in describing features of expansion of the universe.
$q(z)$ is positive (negative) for a decelerating (accelerating) universe. 
Evolution of the jerk
parameter $j(z)$ is relevant in search for for departure from $\Lambda$-CDM model \cite{cg4}.
Exploiting the $d/dt = -(1+z)H(z)d/dz$  in above equations we
in above equations we express $q(z)$ and $j(z)$ as
\begin{eqnarray}
q(z) &=&   -1 + \frac{1}{2}(1+z)\frac{[E(z)^2]^\prime}{E(z)^2} \label{eq:decl}\\
j(z) &=&    \frac{1}{2}(1+z)^2 \frac{[E(z)^2]^{\prime\prime}}{E(z)^2}  \,
- (1+z)\frac{[E(z)^2]^\prime}{E(z)^2} + 1 \label{eq:jerk}
\end{eqnarray} 
where $^\prime$ in above two equations denote derivative with respect to $z$. Using
Eq.\ (\ref{eq:ez}) with $\Omega_{r}^0 = 0$ and $\Omega_{\rm de}^0 = 1 - \Omega_m^0$
in Eqs.\ (\ref{eq:decl}) and (\ref{eq:jerk}) and using expressions for $w(z)$
in terms of parameters $w_a$ and $w_b$ for different models considered in this work
(as summarised in Tab.\ \ref{tab:1}) we may express corresponding
$z-$dependences of $q(z)$ and $j(z)$
involving parameters $w_a$, $w_b$ and $\Omega_m^0$. Putting $z=0$, we may then
obtain relationships between $w_a$, $w_b$, $\Omega_m^0$, $q_0$ and $j_0$,
where $q_0$ and $j_0$ corresponds to the values of deceleration parameter 
and jerk parameter at present epoch. We may take these two parameters ($q_0$ and $j_0$)
as cosmographic parameters relevant in this context. In the methodology
of analysis described in Sec.\ \ref{sec:analysis}, apart from the
nuisance parameters $\alpha$ and $\beta$, the set $(\Omega_m^0, q_0, j_0)$
in stead of the set $(\Omega_m^0, w_a, w_b)$
may be chosen for a cosmographic analysis. 
A comprehensive statistical 
analysis of  specific DE parametrisations using SNe Ia data
have been performed in \cite{cg1}, which directly gives cosmographic
parameters values. For a qualitative comparison, we have given
in Tab.\ \ref{tab:4},
the expressions for $q_0$ and $j_0$ for different DE parametrisations
in terms of $w_a$, $w_b$ and $\Omega_m^0$.
As discussed in Sec.\ \ref{sec:results1},
for each of the varying dark energy Models,
we obtained a point in $w_a - w_b$ parameter space 
(marked by $P_1$ in Fig.\ \ref{fig:2}), 
corresponding to  the parameters satisfying the physical bound $0<c_s^2<1$.
Values of these best-fits $P_1$, are given in the last column of Tab.\ \ref{tab:2}.
For a comparison with the results of comprehensive cosmographic
analysis performed in \cite{cg1}, we have also given in Tab.\ \ref{tab:4}, the numerical
values of $q_0$ and $j_0$ calculated using the their analytical
expressions given in 2nd and 3rd column of the same table, at the point $P_1$.

\begin{table}[t!]
\begin{center}
 \begin{tabular}{l|l|l|l|l|}
\hline
Model & $q_0(w_a,\Omega^0_m)$ & $j_0(w_a,w_b,\Omega^0_m)$ &$q_0$ at $P_1$ & $j_0$ at $P_1$ \\
&&&&\\
\hline
CPL  &  & $\frac{\Omega^0_{de}[3w_b+2+9w_a(w_a+1)]+2\Omega^0_m}{2(\Omega^0_{de}+\Omega^0_m)}$& -0.39 & 1.71\\
\cline{3-5}
JBP  & $ \frac{(3w_a+1)\Omega^0_{de}+\Omega^0_m}{2(\Omega^0_{de}+\Omega^0_m)}$ & $\frac{\Omega^0_{de}[9w_a(w_a+1)+3w_b]+2\Omega^0_m}{2(\Omega^0_{de}+\Omega^0_m)} $& -0.08 & 1.96\\
\cline{3-5}
BA  &   & $\frac{3\Omega^0_{de}[w_b+3w_a(w_a+1)]+2\Omega^0_m}{2(\Omega^0_{de}+\Omega^0_m)} $& -0.52 & 0.58\\
\cline{3-5}
Logarithmic  &  & $\frac{\Omega^0_{de}[3w_a^2+3w_b-3w_a-4]+2\Omega^0_m}{2(\Omega^0_{de}+\Omega^0_m)} $& -0.47 & 1.36\\
\hline
\end{tabular}
\end{center}
\caption{\label{tab:4}  
Expressions of cosmographic parameters 
$q_0$ and $j_0$ in terms of $w_a,w_b,\Omega_m^0$ ($\Omega_{\rm de}^0 = 1 - \Omega_m^0$)
for different  varying dark energy Models.
Numerical values of $q_0$ and $j_0$
calculated using the expressions
at point $P_1$ (see text for details) are also shown.}
\end{table}

\section{Conclusion}
\label{sec:conclusion} 

In this work we have performed a comprehensive analysis of 
recently released `Joint Light-curve Analysis' (JLA) data
to investigate its implications for  models of dark energy
 with varying equation of state parameter $w(z)$.
As a benchmark, we considered 4 different varying dark energy models,
  \textit{viz.} CPL\cite{CPL1}, JBP\cite{JBP1},\cite{JBP2}, BA(\cite{Barboza1},\cite{Barboza2}) 
and Logarithmic model \cite{Sangwan}, each of which involves
a specific functional form of $z-$dependence of the dark energy equation
of state $w(z)$. The analytical expression for the function 
$w(z)$ in each case involves two parameters, denoted by $w_a$ and $w_b$.
From the analysis of observational data we have obtained best-fit values
of these parameters and also their ranges allowed at $1\sigma$, $2\sigma$ and $3\sigma$
confidence level. Description of the data and methodology of analysis has been discussed in 
detail in Sec.\ \ref{sec:analysis}. The results of the analysis are presented in
table\ \ref{tab:2} and depicted in Fig.\ \ref{fig:2}.\\

We make an attempt to realise the scenario of varying equation state of dark energy in terms of 
dynamics of a scalar field $\phi$. We assume the scalar field to be homogeneous  
with its dynamics   driven by a $k-$essence Lagrangian $L = VF(X)$, with a constant potential
$V$ and a dynamical term $F(X)$ with $X=(1/2)\nabla^\mu\phi\nabla_\mu\phi$.
Consideration of constant  potential ensures a scaling relation
of the form $X(dF/dX)^2 = Ca^{-6}$ ($C=$ constant) in FLRW spacetime background with scale factor
$a$. We have exploited this to reconstruct functional form of  $F(X)$
for the different varying dark energy models considered here (See Sec.\ \ref{sec:kessence}
for details). 
The nature of  $F(X)$ are obtained for each kind of $w(z)$ dependences corresponding 
to the best-fit parameters ($w_a,w_b$).  
In this context, we also obtain the dependences of $c_s^2$ on $z$.
$c_s$, as mentioned earlier, is the  the speed with which small fluctuation in the 
background energy density grows. Stability of the density perturbations
and causality requires $c_s^2$ to lie in the domain $0<c_s^2<1$.
The results show  that at the best-fits, $c_s^2$ lies  within its physical bound   
$0<c_s^2<1$ only for a small range of values of $z$. 
For each
of the varying DE models we have shown the region  of the 
$w_a-w_b$ parameter space for which $c_s^2$ lies with its physical bound 
for all values of $z$ accessible in
SNe Ia observations. 
We finally find the point $P_1$ in 
$w_a-w_b$ parameter space  for which $0<c_s^2<1$ 
for all   $z$  and which is maximally favoured from the
observational data. The $z-$dependence of $c_s^2$ and  form of $F(X)$ are 
also obtained for this point ($P_1$) for all the varying DE models.
These results are discussed in detail in Sec.\ \ref{sec:results1} 
and depicted in Fig.\ \ref{fig:3}. In this section we have also given a comparison of 
our results with those obtained from a comprehensive cosmographic analysis
performed in \cite{cg1}.

\section{Acknowledgments}
 We would like to thank the honourable referee for  valuable suggestions.
 AC would like to thank University Grants Commission (UGC), India for supporting this work by
 means of NET Fellowship (Ref. No. 22/06/2014 (i) EU-V and Sr. No. 2061451168).

\end{document}